\documentclass[usenatbib]{mn2e}
\usepackage{graphicx}
\usepackage{subfigure}
\begin{document}

\title{The role of spin in the formation and evolution of galaxies}

\author[Berta et al.]{Zachory K. Berta$^1$, Raul Jimenez$^{1,2}$\thanks{email: raul@ieec.uab.es}, Alan F. Heavens$^3$, Ben Panter$^3$\\
$^1$Institute of Space Sciences (CSIC-IEEC)/ICREA, Campus UAB, Barcelona 08193, Spain.\\
$^2$Dept. of Astrophysical Sciences, Peyton Hall, Princeton University, NJ-08544, USA.\\
$^3$SUPA, Institute for Astronomy, University of Edinburgh, Blackford Hill, Edinburgh EH9 3HJ, UK.}
\date{13 June 2008}

\maketitle

\begin{abstract}
Using the SDSS spectroscopic sample, we estimate the dark matter halo spin parameter $\lambda$ for $\sim53,000$ disk galaxies for which MOPED star formation histories are available. We investigate the relationship between spin and total stellar mass, star formation history, and environment. First, we find a clear anti-correlation between stellar mass and spin, with low mass galaxies generally having high dark matter spins. Second, galaxies which have formed more than $\sim5\%$ of their stars in the last 0.2 Gyr have more broadly distributed and typically higher spins (including a significant fraction with $\lambda > 0.1$) than galaxies which formed a large fraction of their stars more than 10 Gyr ago. Finally, we find little or no correlation between the value of spin of the dark halo and environment as determined both by proximity to a new cluster catalogue and a marked correlation study. This agrees well with the predictions from linear hierarchical torquing theory and numerical simulations. 
\end{abstract}


\section{Introduction}

The influence of tidal torques on the evolution of galaxies was
recognized very early on \citep{Hoyle,Doro70} as a driver of galaxy
morphology. Later on, the spin parameter $\lambda$ of dark matter halos, which was introduced by \citet{Peebles} as

\begin{equation}
\lambda = \frac{J|E|^{1/2}}{GM^{5/2}}
\label{eq-lambda-peebles}
\end{equation}
where $J$ is the angular momentum, $E$ is the energy, and $M$ is the total mass of the galaxy, was studied analytically and numerically within the framework of hierarchical galaxy formation by several authors (e.g. \citet{BE87,HP88,warren92,CT96}). 

%Spin is expected exhibit a roughly log--normal distribution with a broad range of spin parameters ($0.01 \leq \lambda \leq
%0.1$), to increase for decreasing halo mass \citep[]{PEOPLE}, and to correlate weakly with peak height in the primordial density field (and therefore environment) \citep{PEOPLE}.

{ \citet{Vitvitskaetal2002}, \citet{Maller&Dekel2002} and \citet{Hetznecker&burkert} have put forth that in addition to tidal torquing, galaxy mergers play a driving role in the evolution of galactic spins. While they find that recent major mergers increase a halo's spin, \citet{DOnghia&Navarro} emphasize that mergers influence spin only as long as the halo is out of equilibrium, and that isolated, virialized haloes have spins consistent with simple tidal torques. }

It is reasonable to expect that the spin of a dark matter halo will influence the final properties of baryonic matter in the galaxy. For a simple example, consider a disk galaxy and assume baryons settle into the disk with no loss of angular momentum. As given in Eq.~\ref{eq-lambda-peebles}, $\lambda$ measures the degree to which rotation contributes to supporting the galaxy against collapse, between negligibly ($\lambda=0$) and completely ($\lambda \sim 1$). Higher $\lambda$ disks are more rotationally supported and will therefore be less dense. When coupled with a star formation law dependent on density \citep{Kennicutt}, higher $\lambda$ further implies less efficient star forming systems.

Using analytical models, \citet{Dalcanton97}, \citet{JHHP97}, \citet{vdB98}, and \citet{mo} showed in detail how a distribution in the values of the halo spin parameter could lead to
significant differences in the star formation efficiency of the disk, thus shaping the history of the galaxy,
and therefore the morphological type, even leading to possible cases of
dark galaxies \citep{Dalcanton97,JHHP97,VOJ02}, where star formation in the disk has been completely prevented.

It is of great interest to measure observationally the dark halo spin and determine its link star formation, {especially in view of the possible role of mergers on influencing $\lambda$ and their known link to star formation}. In a previous study, \citet{JVO03} used a sample of observed rotation curves of spiral galaxies to  determine the spin distribution $P(\lambda)$, showing that it was consistent with that predicted by hierarchical tidal torque theory (see their Fig.~2). \citet{JVO03} also showed that the spin parameter correlated with the baryon fraction in the galaxy: the higher the spin the higher the baryon fraction. 

Studies of the spin parameter in dark matter numerical simulations have recently reached maturity in that the number of particles per halo is now large enough
for statistical samples of the spin parameter to be reliably measured \cite[e.g]{Gouda98,bailin,avila,gott,maccio,bett, vandenBoschetal2002, Hetznecker&burkert, DOnghia&Navarro}. Numerical simulations seem to agree with the early analytical calculations, confirming that $\lambda$'s dependence on environment is very weak \citep{falten}.

Problematically, the spin parameter cannot be measured easily from observations, as none of $J$, $E$ or $M$ can be measured directly. As mentioned above, \citet{JVO03} used an indirect method to determine the halo spin parameter, where they assumed a particular dark matter halo profile and an exponential disk for the baryonic matter and fit a parameterized model to observed rotation curves to recover $\lambda$. Recently, \citet{HC-S06} (hereafter HC-S06) proposed another indirect method for approximating $\lambda$, in the same vein as \citet{JVO03} but without needing to use rotation curve information.

Like \citet{JVO03}, HC-S06 assume the specific angular momentum of the dark matter and the baryons are equal, but they determine the galactic rotation velocities with the Tully-Fisher relation. They show from comparison with numerical simulations that
their estimator gives an accurate measurement of $\lambda$ \citep{Hernandez08}. \citet
{Hernandez07} applied this to the SDSS survey to determine the spin
parameters of a volume limited sample of $~8,000$ color--and concentration--selected SDSS spiral galaxies, finding that the
spin parameter distribution $P(\lambda)$ does indeed follow a log-normal distribution in very good agreement with that predicted by
analytic or numerical calculations. In a more recent study \citep{Hernandez08}, the same group computed the correlation of $\lambda$ as a function of mass and of over-density of galaxies in the SDSS, finding an anti-correlation between $\lambda$ and galaxy mass and no correlation between $\lambda$ and environment.

In the present study, by taking advantage of new results and the recent MOPED analysis that provides detailed star formation histories for galaxies in the SDSS-DR3, we modify and improve the HC-S06 method to measure $\lambda$ for a sample of $~53,000$ galaxies. With MOPED we can use stellar masses derived directly from the spectra and thus avoid using the Tully-Fisher law twice as is done in HC-S06. We use our sample to study how the dark matter spin parameter shapes the star formation history of galaxies. We also investigate how the spin depends on environment as measured by both the galaxies' distances to catalogued galaxy clusters and a $\lambda$-weighted marked correlation analysis for comparison with the \citet{falten} numerical simulations. We find that active star forming galaxies today have a significantly broader spin distribution, reaching to significantly higher values of $\lambda$, than galaxies that formed most of their stars at $z > 2.5$. We also show that spin shows little dependence on environment, thus confirming the analytical and numerical results from hierarchical torquing theory. Our paper is organized as follows: in \S 2 we describe the sample of galaxies used and in \S 3 the method for approximating $\lambda$.  We present our main results in \S 4 and our conclusions in \S 5.

\section{Sample}

Our original sample contains $\sim3\times10^{5}$ galaxies for which MOPED star formation histories have been computed \citep{Panter07}. The galaxies were selected from the SDSS-DR3 Main Galaxy Sample \citep{Strauss02, DR3}, satisfying $15.0\leq r \leq 17.77$ and spanning the redshifts $0.01\leq z \leq 0.25$. For each galaxy, MOPED computes the star formation fraction (SFF) in 11 roughly logarithmically spaced time bins, the metallicity in each of the 11 bins, and the estimated $E(B-V)$ dust extinction (which includes both internal and Galactic dust). MOPED also estimates the present stellar mass $M_{*}$ of each galaxy, along with the total mass of stars that has ever formed in that galaxy. More about the MOPED algorithm and sample can be found in \citet{Panter07}.

Of those galaxies with known star formation histories, our sample includes only disk-dominated galaxies (SDSS fractional de Vaucouleurs weight {\tt fracDeV < 0.5}) which are viewed face-on (SDSS observed axis ratio {\tt expAB > 0.6}). While this latter cut is rather generous, MOPED's dust estimate empowers a good correction for internal extinction. Unless otherwise stated, the results presented below persist for more stringent cuts on both of these parameters, albeit with higher noise. This subset contains $\sim5.3\times10^{4}$ galaxies. {Where necessary, photometric parameters from DR6 are used \citep{DR6}.}
\section{Method}

Since none of the quantities in Eq.~\ref{eq-lambda-peebles} is directly measurable, some estimation is required. The more accurate method to determine $\lambda$ would be to employ galaxy rotation curves along with detailed modeling of galactic potentials \citep{JHHP97, Gnedin}, but constructing very large samples in this way is prohibitively difficult. HC-S06 have derived a rough estimator for $\lambda$ in terms of SDSS measurements alone; it relies on the Tully-Fisher relation \citep{TF} to provide disk velocities where necessary.

We will give a quick overview of the HC-S06 method and will note where we modify their method to obtain more robust $\lambda$s.
The HC-S06 model contains a baryonic galactic disk of mass $M_{d}$ with an exponential profile
\begin{equation}
\Sigma (r) = \left(\frac{M_{d}}{2\pi R_{d}^2} \right)e^{-r/R_d},
\end{equation}
where $r$ is the radial coordinate and $R_{d}$ the scale radius. This disk sits in a truncated, spherical, isothermal halo of mass $M_{h}$ which gives it a circular velocity $V_{d}$. In this simplified model, the halo dominates the quantities $J$, $E$, and $M$; the disk is an observable tracer of them but contributes negligibly to their totals. Our data limitations prevent us from any more detailed modeling of the dark halo NFW profile, but this toy model should be a good approximation, as long as galaxies have flat rotation curves.

\begin{figure}
\includegraphics[width=\columnwidth]{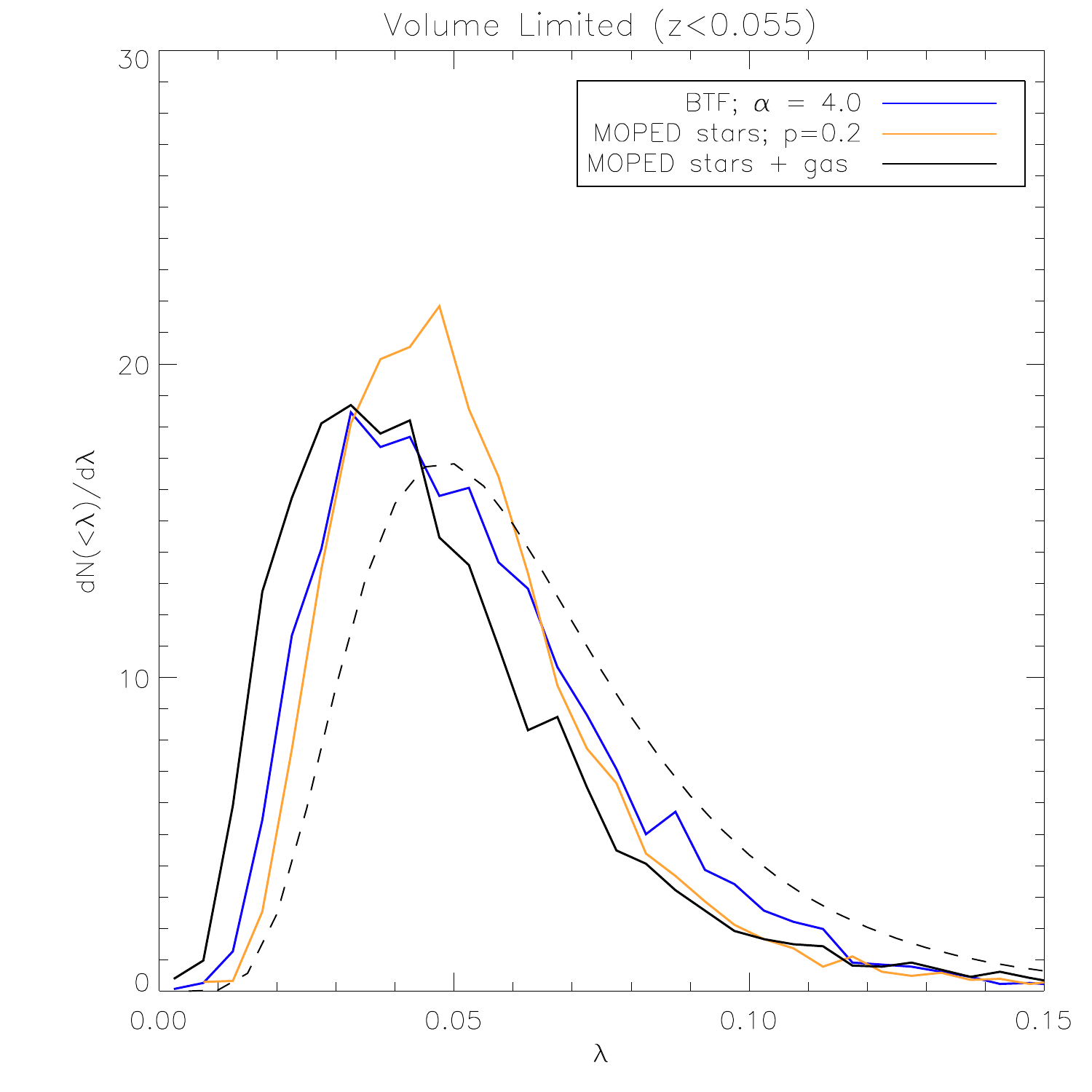}
\caption{$P(\lambda)$ for a volume limited sample with $z<0.055$ and $M_{r}<-19.0$ (chosen for ease of comparison with \citet{HC-S06}) for several estimates of $M_{h}$ (solid colored lines) with the \citet{HC-S06} best fit overplotted (dashed line). }
\label{fig:spin_local}
\end{figure}

To determine $J$, the halo specific angular momentum $j_{h} = J/M_{h}$ is taken to be equal to that of the exponential disk $j_{d} = 2 V_{d} R_{d}$. This is a very good approximation for the protogalactic state, where baryonic and dark angular momenta were well mixed, and is often assumed to hold for collapsed galaxies, as long as their formation histories were smooth \citep{Dalcanton97}, but its robustness is limited if baryons lose a significant fraction of their angular momentum  to dark matter during collapse. {This question is still an open one. Early simulations \citep{NavarroBenz} indicated that the baryons retain only a small fraction of their specific angular momentum through collapse, but \citet{vandenBosch2001} found baryonic spins to relate closely to dark matter spins, albeit with a considerable angular misalignment between the two.} Recent studies (\cite{Kaufmann07,Mayer}) claim that the problem is fraught with numerical difficulties, and that with sufficient resolution, the baryon component retains 80 or 90 percent of its specific angular momentum. Furthermore, physical effects such as supernova feedback effects may play a role.  {In spite of this uncertainty, it} has been common in the literature (e.g. \cite{mo,JHHP97}) to assume that the specific angular momentum of the baryons is the same as that of the dark matter. Assuming the halo to be virialized, $E$ is simply half the gravitational potential energy of the halo: $E = V_{d}^{2} M_{h}/2$.

{If this spin estimator accurately probes the dark matter, it should yield the redefined spin parameter $\lambda'$ introduced by \citet{Bullocketal2001}, as a measure of spin that is independent of density profile. The mapping between $\lambda'$ and $\lambda$ depends on halo shape, and can be difficult in crowded systems with poorly defined edges. Any disparity between the two is well within the expected scatter of the spin estimator, so we will refer to what follows, roughly, simply as $\lambda$.}

In HC-S06, $M_{d}$ is determined from a general baryonic Tully-Fisher relation (BTF),
\begin{equation}
M_{d} = A_{TF} V_{d}^{\alpha}
\end{equation}
where the constants $\alpha = 3.5$ and $A_{TF} = 633 M_{\odot} (km s^{-1})^{-3.5}$. Such an $A_{TF}$ implies a disk mass for the Milky Way ($V_{d} = 220 km/s$) of $M_{d} = 10^{11} M_{\odot}$.

{A critical issue} is that one needs to estimate the mass of the halo in order to estimate $\lambda$.  The best way to do this is not clear, so the procedure we have adopted is to investigate three separate plausible relations between disk mass and halo mass.  We find that all three give very similar quantitative results, and qualitatively the conclusions are identical whichever method is adopted.

HC-S06 assume $F=M_{d}/M_{h}$ is a constant, which they place at $1/25$, and estimate $M_{h}$ for their galaxies. Nominally, the BTF should probe all the baryonic mass in the disk, but since $V_{d}$ is measured only indirectly through another stellar mass TF relation, it introduces a level of abstraction, with stellar light giving rotation velocity which in turn gives baryonic mass.

Since MOPED provides a robust estimate of the disk stellar mass $M_{*}$, we can circumvent the need to rely on the BTF scaling relation and instead use more information from the real, observed galaxies. For high mass galaxies, $M_{*} \approx M_{d}$, but since our sample includes a range of masses, we need to map $M_{*}$ to $M_{d}$ in a consistent manner, taking into account the increasing gas fraction towards lower $M_{d}$. We explore two methods to perform this mapping.

First, we assume the same constant baryon-to-halo fraction $F$, but calculate $M_{d} = M_{*} + M_{g}$ from the MOPED stellar mass $M_{*}$ and an estimate for the associated gas mass $M_{g}$. \citet{Calura07} used the MOPED galaxies and Kennicutt star formation rate scalings to estimate the mean gas fraction $\langle M_{g}/M_{*} \rangle$ as a function of $M_{*}$. We use the results of their Fig. 2 to calculate $M_{g}$.  Note that \citet{Calura07} found some dependence on the fraction of baryons lost as a function of mass (so $F$ is not constant).  However, over the mass range of the vast bulk of the galaxies considered here, the dependence is not large, $\pm 10\%$, and we ignore the variation here.  

Second, following \citet{Gnedin}, we make no gas correction to $M_{*}$ but assume that $M_{h} = M_{*}/F_{*}$ where $F_{*}$ is allowed to vary with galaxy parameters. \citet{Gnedin} suggest that $F_{*}$ (which they call $m_{d}$) should scale with stellar surface density as

\begin{equation}
F_{*} = F_{*,0} \left( \frac{M_{*} R_{d}^{-2}}{10^{9.2} M_{\odot} kpc^{-2} }\right)^{p}
\end{equation}
for some power $p$. As they observe, such a dependence on surface density follows naturally from a Kennicutt-like density-driven star formation law. By comparing observed and modeled TF and Fundamental Plane parameters and residuals, \citet{Gnedin} find a best fit $p=0.2$.

To summarise, to test our assumptions, we examine three assumptions for $M_{h}$: (1) HC-S06's baryonic Tully-Fisher estimate (``{\tt BTF}'') where $M_{h} \approx A_{TF} V_{d}^\alpha/F$, (2) using the MOPED stellar mass with the \citet{Gnedin} variable disk fraction (``{\tt MOPED stars}'') where $M_{h}\approx M_{*}/F_{*}(M_{*})$, and (3) using the gas-corrected MOPED stellar mass (``{\tt MOPED stars + gas}'') where $M_{h} \approx (M_{*} + M_{g})/F$. In all cases, we choose the constants such that the Milky Way has $M_{d} = 10^{11} M_{\odot}$ and $M_{h} = 2.5\times10^{12} M_{\odot}$. For the first case, we use $\alpha = 4.0$ in close agreement with \citet{McGaugh04}, although the results are similar to the HC-S06 choice of $\alpha = 3.5$.

Finally, for each of these cases, we have an approximation for the spin parameter $\lambda$ which looks like \begin{equation}
\lambda \approx \frac{2^{1/2} V_{d}^2 R_{d}}{GM_{h}}.
\label{eq-lambda}
\end{equation}
Here, $R_{d}$ is taken from the SDSS $i$-band {\tt expRad}\footnote{In its exponential fitting, SDSS reports half-light radii $R_{e}$ which relate to our desired radii by $R_{d} = R_{e}/1.68$.}, and the disk velocity $V_{d}$ is determined by the $i$-band Tully-Fisher relation that $\log V_{d} = -0.130(M_{i} + 21.327) +2.212$ published by \citet{Pizagno07}\footnote{They actually publish $V_{80}$, the velocity at a radius encompassing $80\%$ of the $i$-band flux, which ought to be a good estimator of $V_{d}$.}. $M_{h}$ is given by one of the three cases above.  We choose $i$-band for $R_{d}$ and $V_{d}$ to limit potential confusion from active bursts of star formation, and use the MOPED dust parameter to correct for internal and Galactic extinction. \citet{Hernandez07} used an older $R$-band TF relation, which had to be converted to Sloan photometry.

This estimator of $\lambda$ is {\it very} rough, and should not be trusted as precise for any one real galaxy. Typical errors, dominated mostly by scatter in the Tully-Fisher relationships, are of order at least $30\%$ \citep{Hernandez07}. With large numbers of galaxies, we still hope to build reliable $\lambda$ distributions by averaging over the noise. More perniciously, however, unknown systematic offsets in the coefficients severely limit any statements that can be made about absolute values $\lambda$. In deference to these limitations, our conclusions are drawn only from relative comparisons, and should be robust as long as the scalings hold.

{\citet{Hernandez08} test their version of this estimator against previously published N-body simulations, showing that it works reasonably well. Another ideal test would be to calculate detailed spin parameters for a large number of galaxies for which high-resoluation rotation curves are available, following \citet{JVO03, Gnedin}, comparing the results to the estimate based on SDSS parameters. \citet{Haynes} recently published a large sample of such rotation curves which that overlap significantly with the SDSS, but they contain very few galaxies at $z>0.055$, where the {\tt MOPED} analysis can be performed. The cumulative distribution of spins below do agree well with the results of \citet{JVO03}, but no truly solid test of these spin estimations has been performed so they must be interpreted with appropriate caution.}

\begin{figure}
\includegraphics[width=\columnwidth]{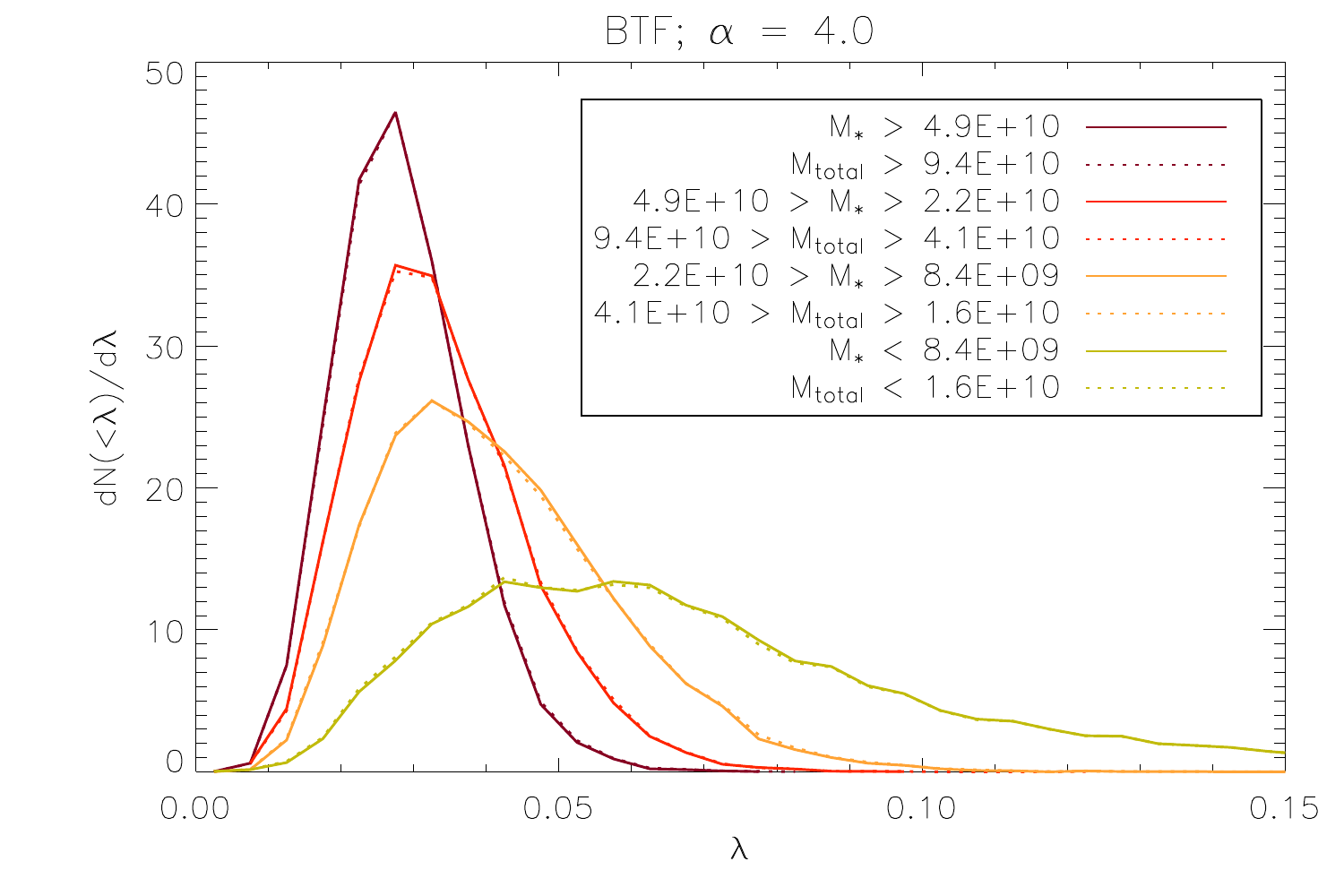}
\includegraphics[width=\columnwidth]{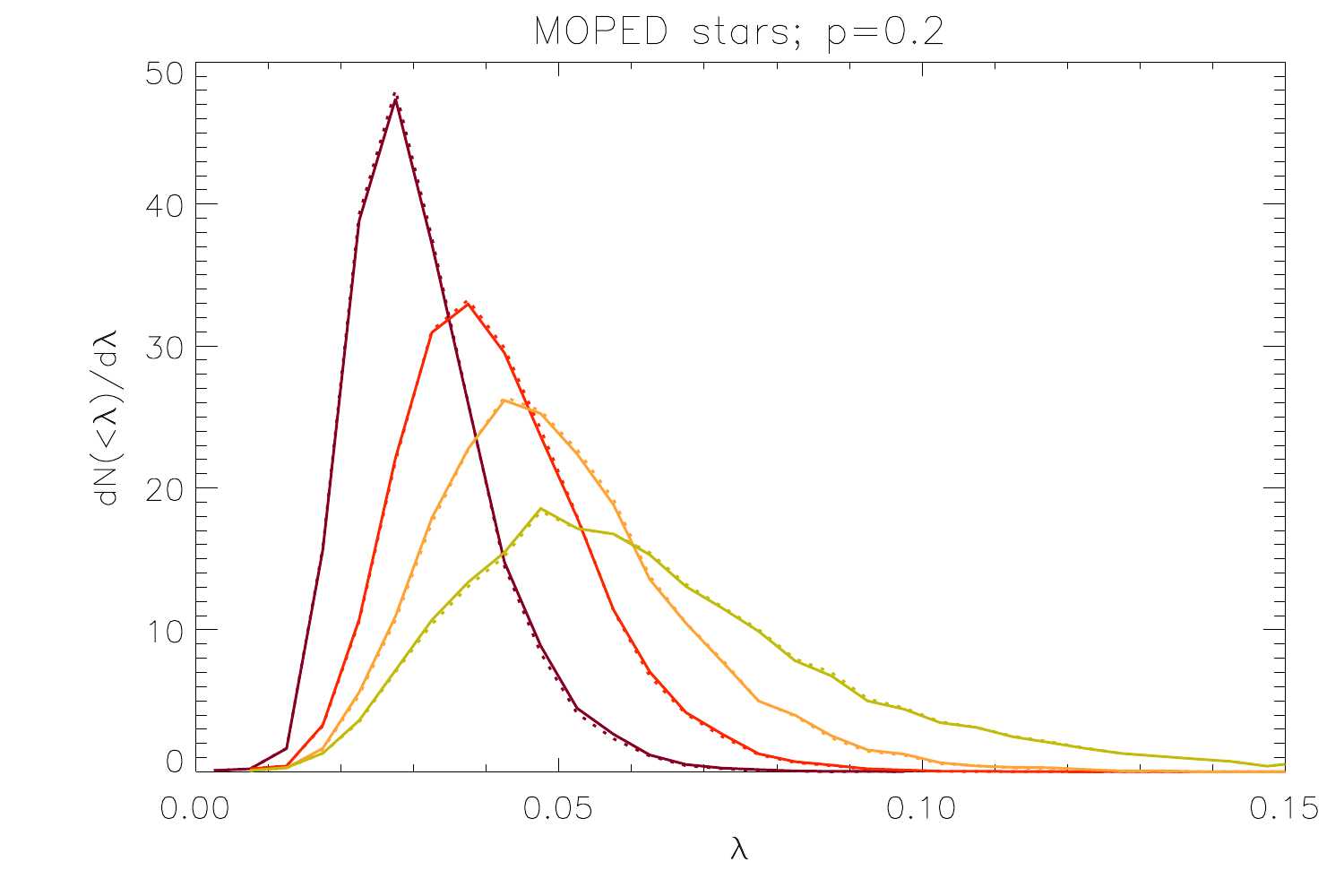}
\includegraphics[width=\columnwidth]{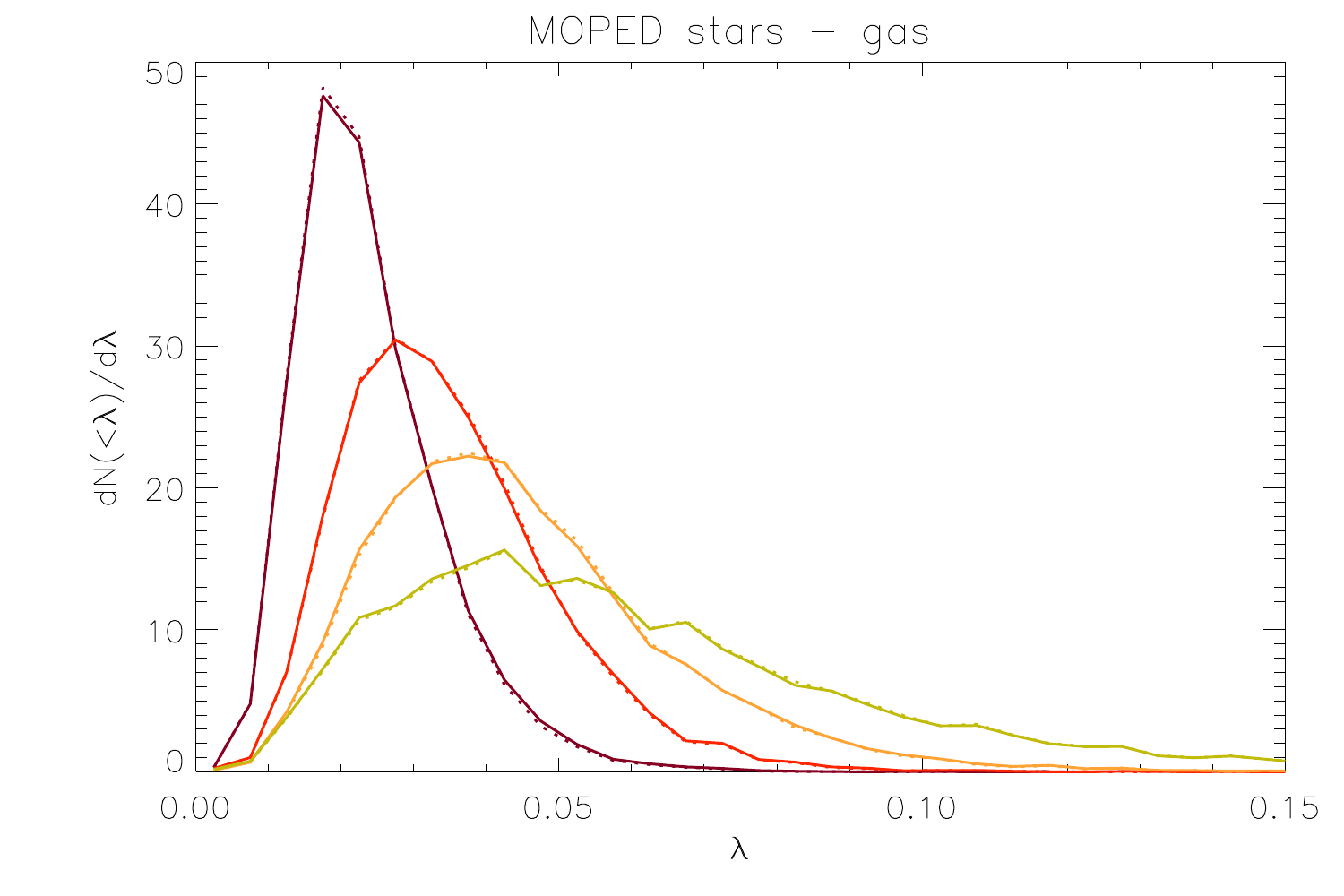}

\caption{For each estimate of $M_{h}$ ({\it top} - {\tt BTF}, {\it middle} - {\tt MOPED stars}, {\it bottom} - {\tt MOPED stars + gas}), $P(\lambda)$ distributions for the entire sample split into quartiles by the present mass of stars in the galaxy $M_{*}$ ({\it solid lines}) and the total mass of stars ever formed in the galaxy $M_{total}$ ({\it dotted lines}). The curves are nearly identical, as would be expected if $M_*$ rises monatonically with $M_{total}$, and the respective mass cuts are shown in the legend in the top panel.}
\label{fig:spin_with_mass}
\end{figure}

\section{Results}

{First, we have tested that we produce a reasonable distribution of $\lambda$'s. Dark matter halos have long been predicted to exhibit a log-normal $P(\lambda)$, with median $\lambda$ ranging from $ \sim 0.03$ to $\sim 0.05$ and logarithmic width $\sigma_{\log\lambda} \approx 0.5 $ \citep{HP88,Vitvitskaetal2002, vandenBoschetal2002, Hetznecker&burkert}.}
 
Our $P(\lambda)$ is plotted in Figure \ref{fig:spin_local} for the several $M_{h}$ estimates discussed above. For all assumptions, the distributions are roughly lognormal and in general agreement with theoretical predictions and the HC-S06 best fit, which is overplotted. Remember that the coefficients going into Eq. \ref{eq-lambda} were independently set to reproduce the mass properties of the Milky Way, and although unknown biases in the estimator could shift these distributions in either direction, they agree quite well.

Next, we investigate how halo spin depends on MOPED stellar mass and star formation history. Fig.~\ref{fig:spin_with_mass} shows $P(\lambda)$ for different stellar masses. Note that the distributions become narrower and with smaller mean as the stellar mass of the galaxy increases. For stellar masses smaller than $1.5 \times 10^{10}$ M$_{\odot}$ the distribution extends to values larger than $\lambda = 0.1$, while for higher masses the values of $\lambda$ end at $\lambda \sim 0.05$. This trend is seen for all our $M_{h}$ estimates. In the {\tt MOPED stars} case, increasing the surface density exponent $p$ effectively decreases $\lambda$ for low mass galaxies, shifting their distribution to the left, and could potentially eliminate this mass dependence. The observed trend, however, persists for $p<0.5$, which seems to be safely far above the \citet{Gnedin} best fit $p=0.2$. This observed anticorrelation between mass and $\lambda$ was also recently observed by \citet{Hernandez08}.

\begin{figure}
\includegraphics[width=\columnwidth]{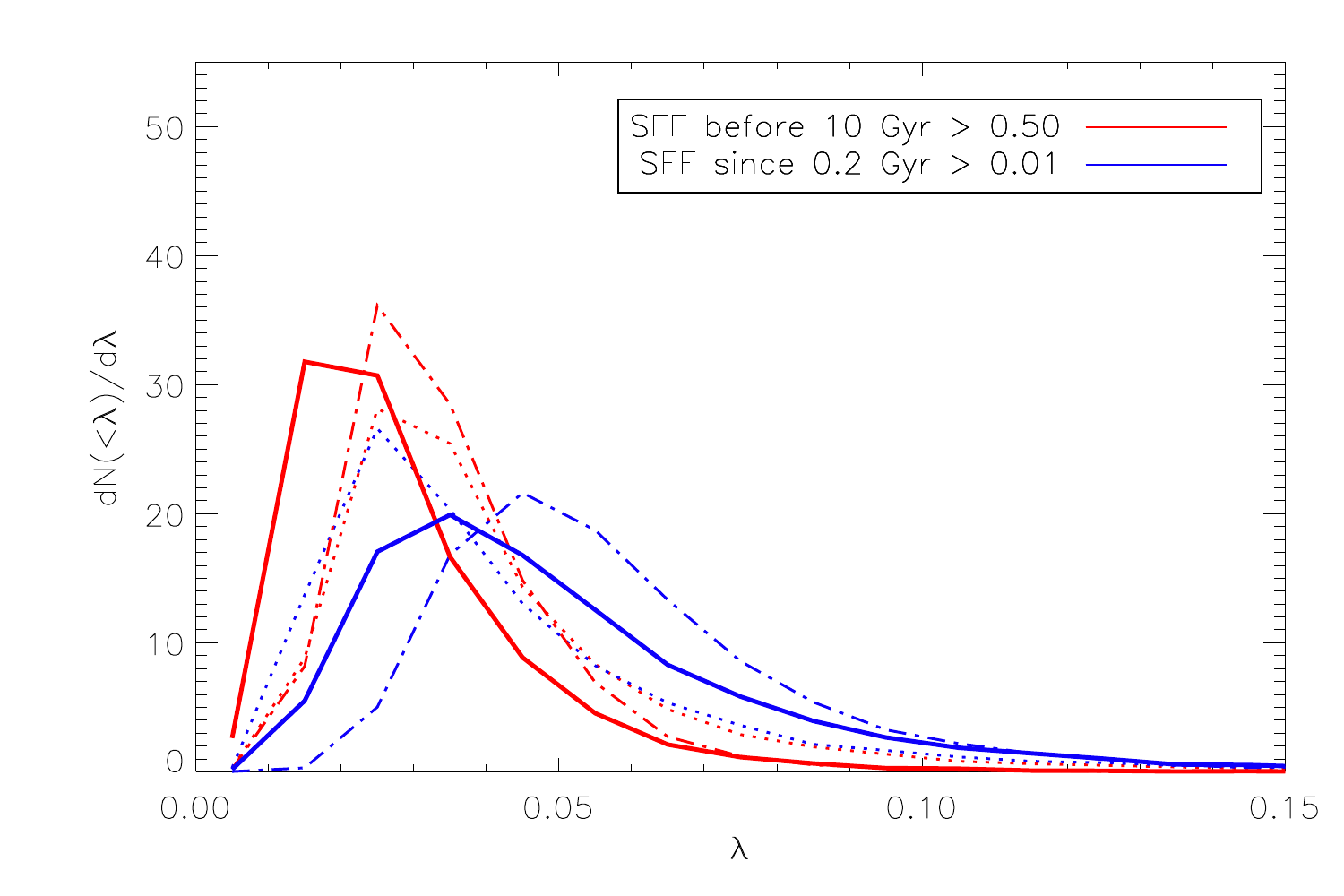}
\includegraphics[width=\columnwidth]{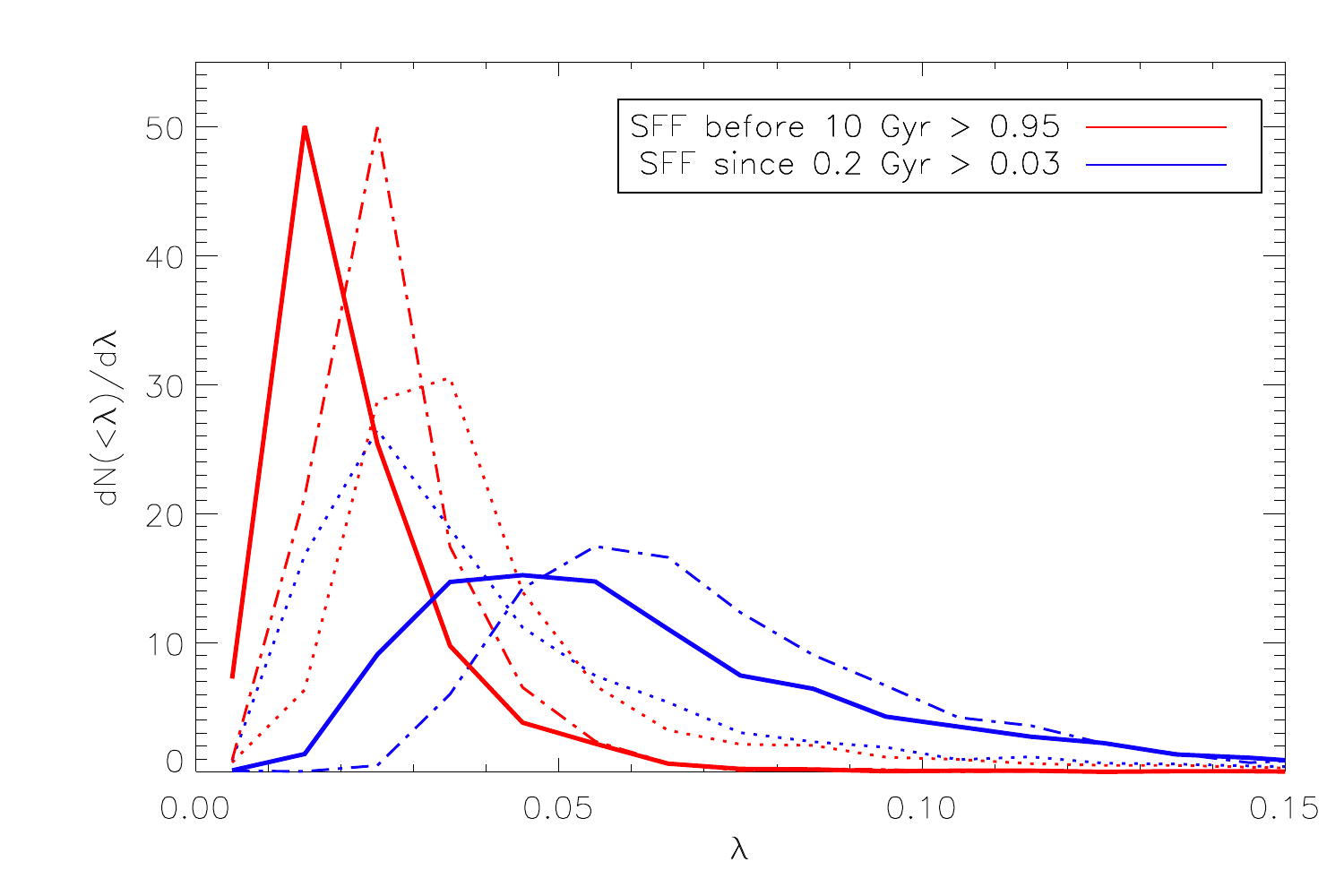}
\includegraphics[width=\columnwidth]{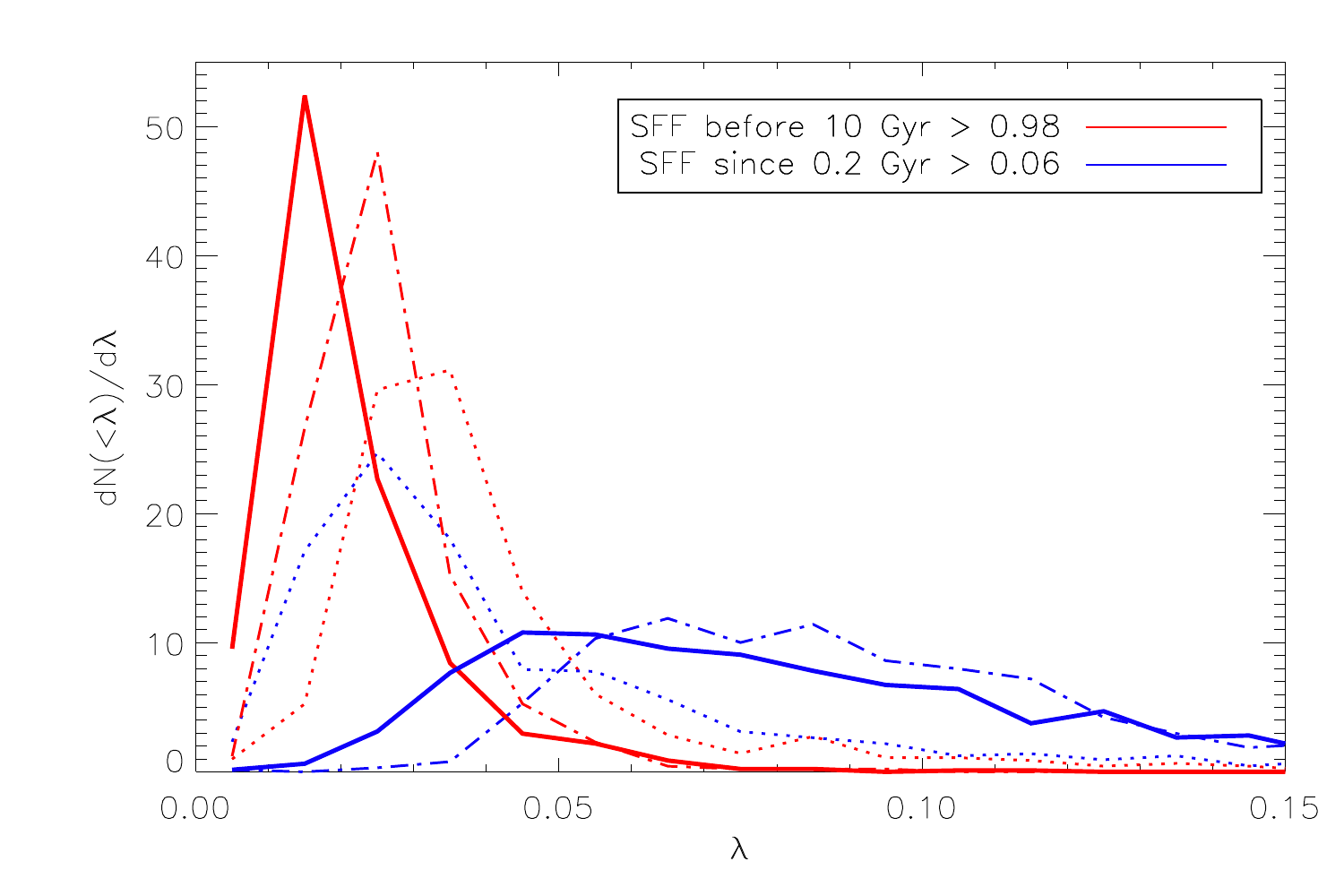}

\caption{For each estimate of $M_{h}$ ({\it dotted lines} - {\tt BTF}, {\it dashed lines} - {\tt MOPED stars}, {\it solid lines} - {\tt MOPED stars + gas}), $P(\lambda)$ is shown for subsamples split on the basis of the fraction of stars formed before 10 Gyr in lookback time ({\it red lines}) and the fraction of stars formed since 0.2 Gyr in lookback time ({\it blue lines}). The star formation fraction (SFF) cuts are shown in each panel, and lower panels show greater extremes of the population. From top to bottom, each curve is calculated from roughly $15,000$, $2,500$, and $1,000$ galaxies respectively. }
\label{fig:sfspin}
\end{figure}

Taking advantage of the previous MOPED analysis, we show in Fig.~\ref{fig:sfspin} how the spin distribution changes as a function of the star formation history. The first panel compares $P(\lambda)$ for galaxies which have formed more than 50\% of their stars at look-back times older than 10 Gyr to that for galaxies which have formed more than 1\% of their stars in their last 0.2 Gyr. The next panel makes more extreme cuts for the star formation fraction in the same time bins (more than 95\% before 10 Gyr and 3\% in the last 0.2 Gyr), and the next even more extreme (98\% and 6\% respectively). For completeness, we show the three $M_{h}$ estimates in each panel of Fig.~\ref{fig:sfspin}. Among both the {\tt MOPED stars} and {\tt MOPED stars + gas} curves, a very clear split is seen between narrowly peaked low-$\lambda$, old stellar population galaxies and broadly distributed high-$\lambda$, young stellar population galaxies. Furthermore, as the definitions of these regimes are made more extreme (higher SFF, lower panels), the differences between the $\lambda$ populations widen. {The {\tt BTF} curves are slightly more enigmatic, with the peaks shifting in the opposite direction, but with the {\tt MOPED} trend persisting out in the tails.} Since the {\tt MOPED stars} and {\tt MOPED stars + gas} methods more directly probe the true mass of each galaxy we suggest those as representative of the underlying spin distributions.

The recent star formation populations in Fig.~\ref{fig:sfspin} do appear to roughly overlap with the low-mass quartiles in Fig.~\ref{fig:spin_with_mass}, in accordance with the finding by \citet{Heavens04,Panter07} that galaxies with total stellar masses below $5\times 10^{10}$ M$_{\odot}$ dominate star formation today. {If star-forming galaxies are typically low mass, it appears from Fig~\ref{fig:sfspin} that they are also generally high spin.} But furthermore, $30-60\%$ of the galaxies contributing to the three recent star formation curves in Fig.~\ref{fig:sfspin} have total stellar masses above $8\times 10^{9}$ M$_{\odot}$, suggesting that star formation is occurring today in high spin galaxies of all masses.

\begin{figure*}[
\begin{minipage}[b]{0.9\columnwidth}
\centering
\includegraphics[width=0.9\columnwidth]{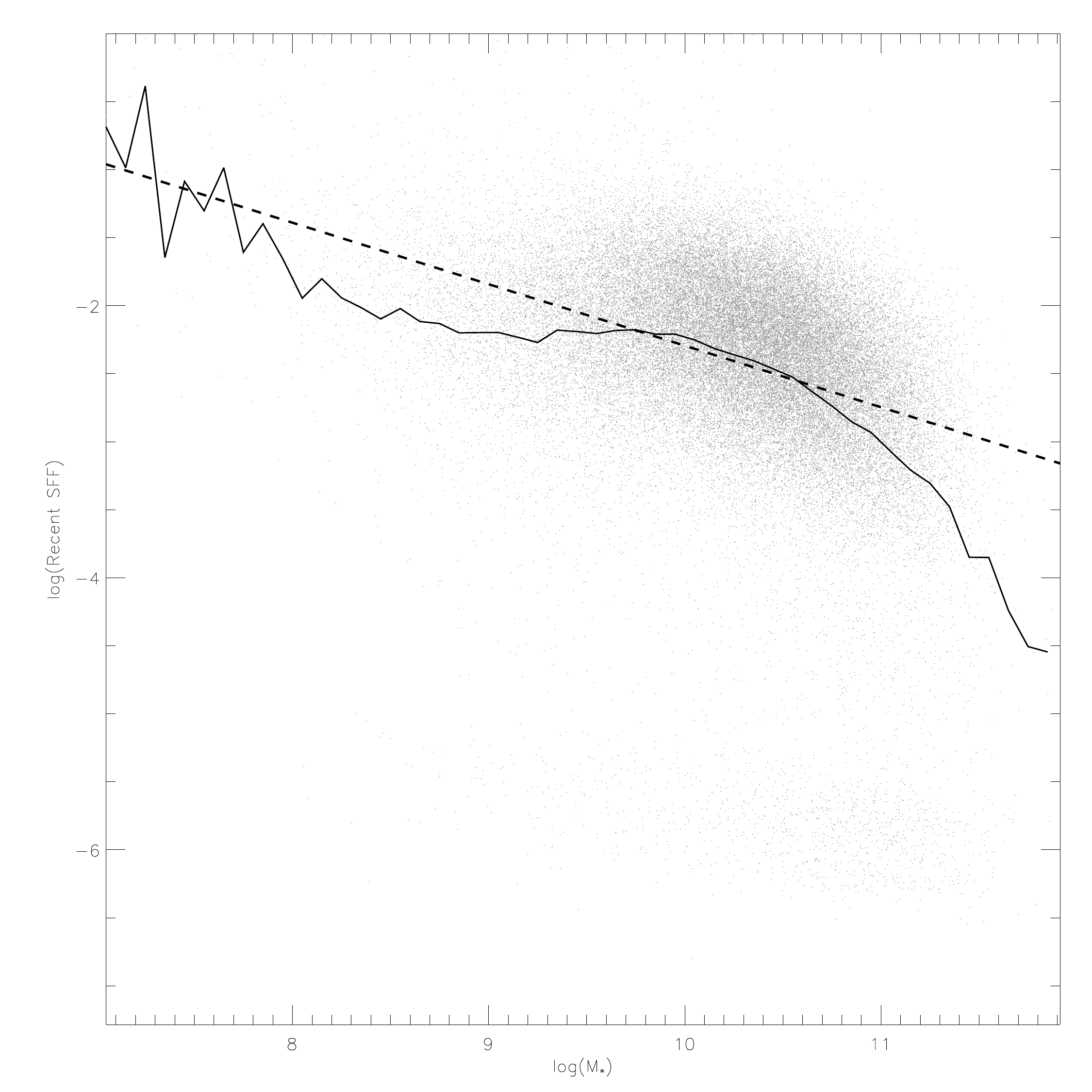}
\end{minipage}
\hspace{0.5cm}
\begin{minipage}[b]{0.9\columnwidth}
\centering
\includegraphics[width=.9\columnwidth]{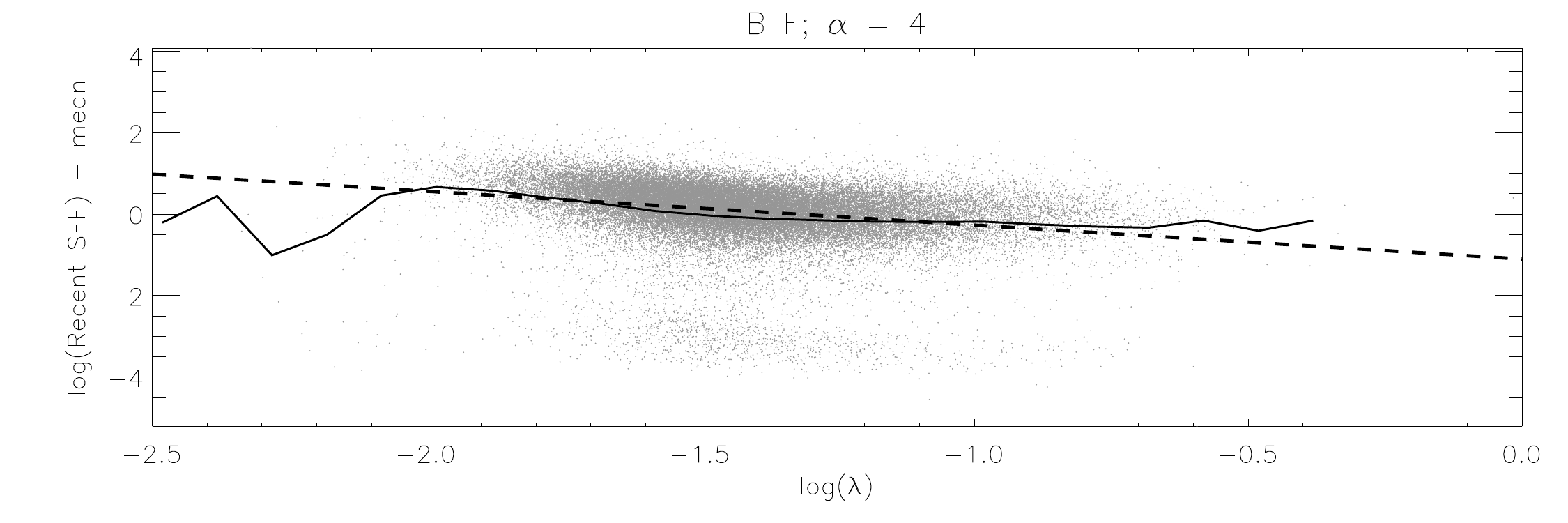} \\ 
\includegraphics[width=.9\columnwidth]{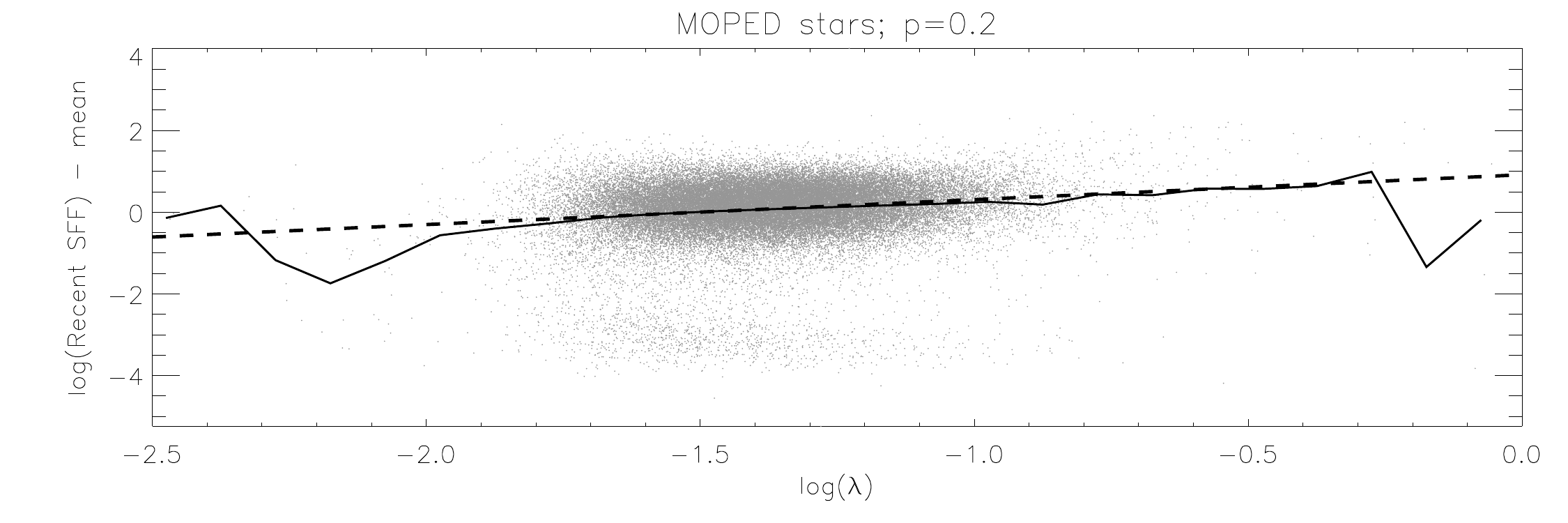} \\ 
\includegraphics[width=.9\columnwidth]{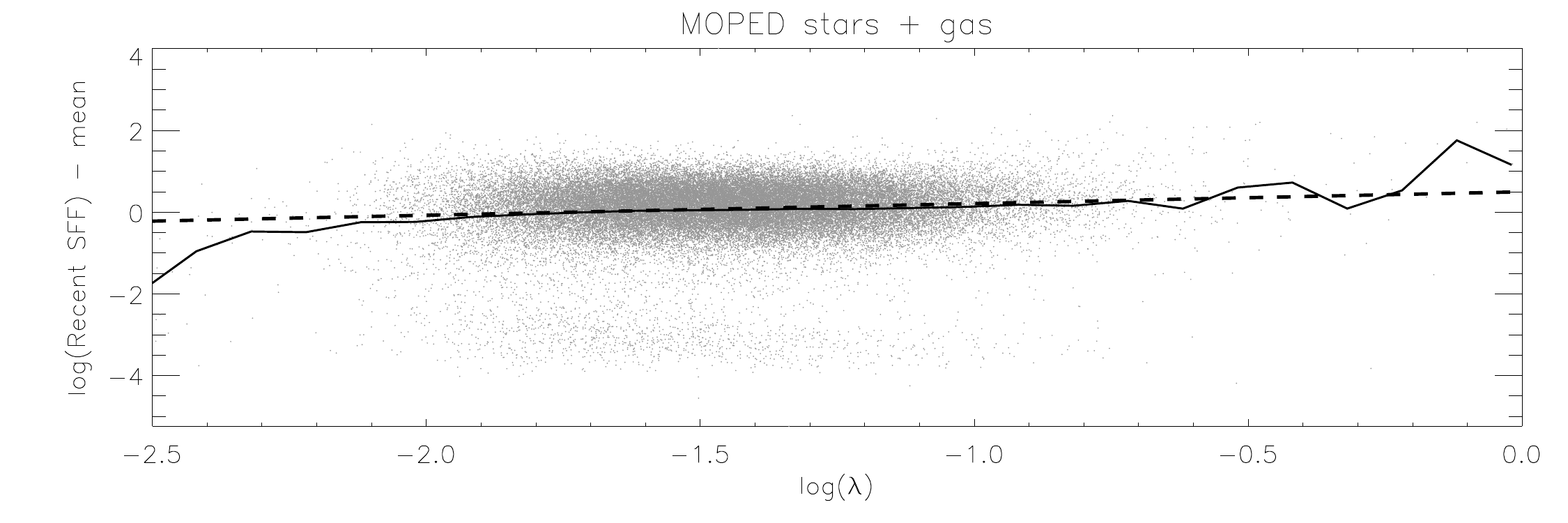}
\end{minipage}
\caption{{\it Left:} Scatter plot showing the recent star formation ($<$ 0.2 Gyr) as a function of stellar mass $M_{*}$, with the best linear fit ({\it dashed line}, both panels) and the mean in bins of $\Delta \log M_{*}=0.1$ ({\it solid line}, both panels). {\it Right:} The residual, having subtracted the binned mean ({\it solid line}) from each point at left, as a function of spin for three halo mass estimates: {\tt BTF} ({\it top}), {\tt MOPED stars} ({\it middle}), {\tt MOPED stars + gas} ({\it bottom}). }
\label{fig-res_fit}
\end{figure*}

{Fig.~\ref{fig-res_fit} tries to pinpoint whether the correlation between spin and star formation seen in Fig.~\ref{fig:sfspin} is due primarily to link between mass and star formation or if spin provides a second parameter for star formation history. We define the ``Recent SFF'' as the fraction of stars formed within the past 0.2 Gyr, and plot it against stellar mass $M_{*}$ in Fig.~\ref{fig-res_fit}. As discussed extensively in \citet{Panter07}, it shows that recent star formation occurs predominantly in low-mass galaxies. To determine if spin plays an additional observable role in galactic star formation, in the right panel we plot the residuals from the binned mean of the left plot as a function of spin for each of the three $M_{h}$ estimates. The best fit lines for the residuals, of the form

\begin{equation}
\log (\mbox{Recent SFF}) - mean(M_{*}) = a \log (\lambda)+b,
\label{eq:plane}
\end{equation}
are $a=-1.1, b=-0.8$ with a scatter of $\sigma_{\mathrm{Recent~SFF}} = 0.47$ ({\tt BTF}), $a=0.92, b=0.61$ with $\sigma_{\mathrm{Recent~SFF}} = 0.49$ ({\tt MOPED stars}),  and $a=0.50, b=0.29$ with $\sigma_{\mathrm{Recent~SFF}} = 0.50$ ({\tt MOPED stars + gas}). That the {\tt BTF} residual ({\it top}) shows a negative slope is consistent with the enigmatic result seen in Fig.~\ref{fig:sfspin}, where by any measure the correlation between $\lambda$ and SFF is less clear. Both the {\tt MOPED stars} ({\it middle}) and {\tt MOPED stars + gas} ({\it bottom}) show slight positive slopes, indicating that high-$\lambda$ galaxies for a given $M_{*}$ host an excess of recent star formation. Although the scatter is considerable, Pearson's $\rho$ has a virtually zero null probability in each case.

The discrepancy between the three $\lambda$ estimates is troubling. Since we expect that the {\tt MOPED} masses more accurately reflect true galaxy masses, as they are not plagued by the {\tt BTF}'s double use of the same scaling relation, Fig.~\ref{fig-res_fit} could indicate a slight dependence on spin as a second parameter governing the star formation history of galaxies. However, because ultimately we sample only the luminous disk matter whose link to halo properties may be very complicated, the question of spin an independent drivers of star formation history remains poorly resolved.}
%We also fit a plane in the $\lambda$-$M_{*}$-(Recent SFF) space, of the form
%\begin{equation}
%\log (\mbox{Recent SFF}) = c \log \left(\frac{{\rm M}_{*}}{\rm M_\odot}\right) + d \log (\lambda)+e,
%\label{eq:plane}
%\end{equation}
%with $c=-0.84, d=-1.55, e=3.98$ with a scatter of $\sigma_{\mathrm{Recent~}} = 0.47$ ({\tt BTF}), $c=-0.22, d=1.22, e=1.52$ with $\sigma_{\mathrm{Recent~SFF}} = 0.50$ ({\tt MOPED stars}),  and $c=-0.31, d=0.63, e=1.65$ with  $\sigma_{\mathrm{Recent~SFF}} = 0.47$ ({\tt MOPED stars + gas}).

\begin{figure*}
\includegraphics[width=0.68\columnwidth]{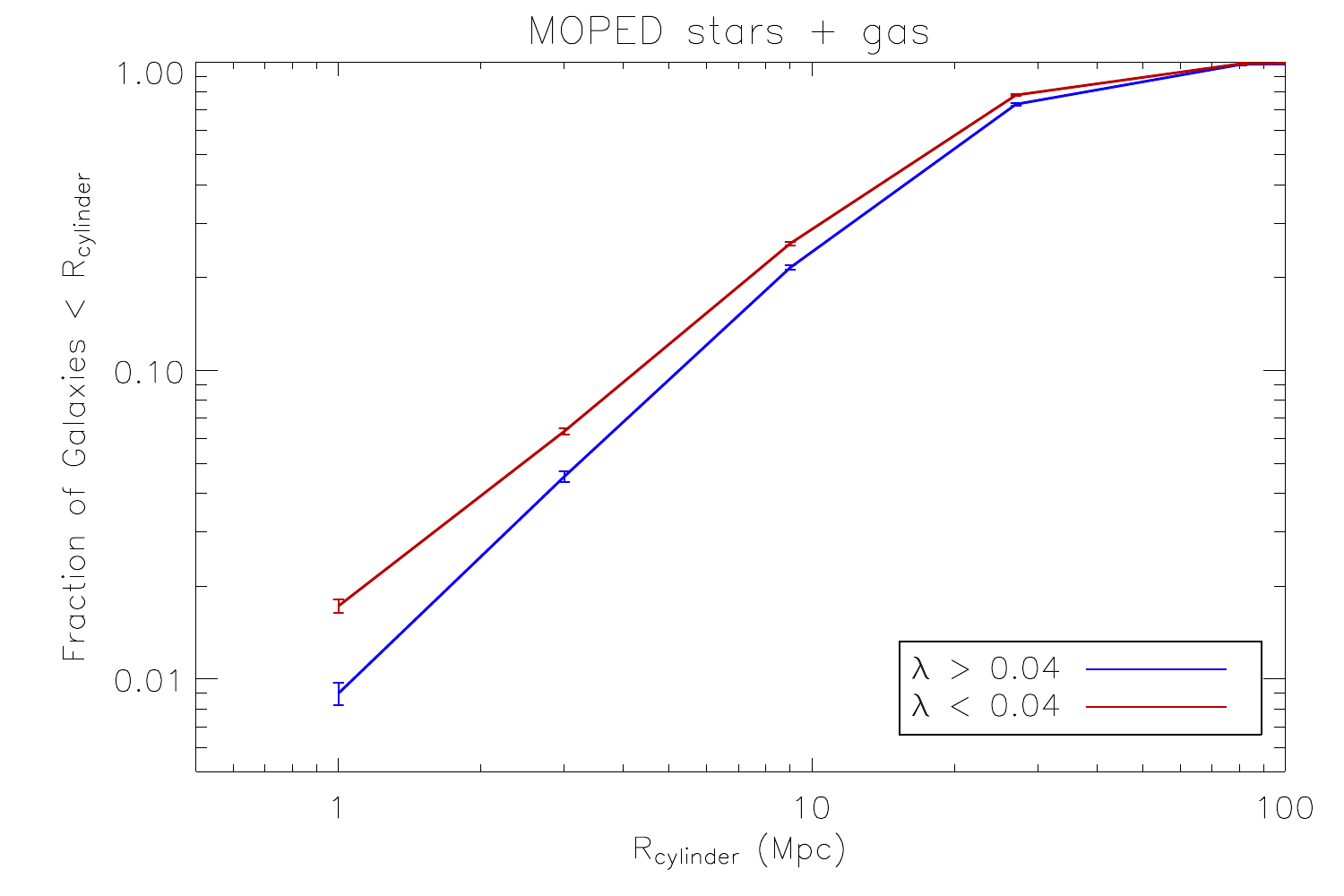}
\includegraphics[width=0.68\columnwidth]{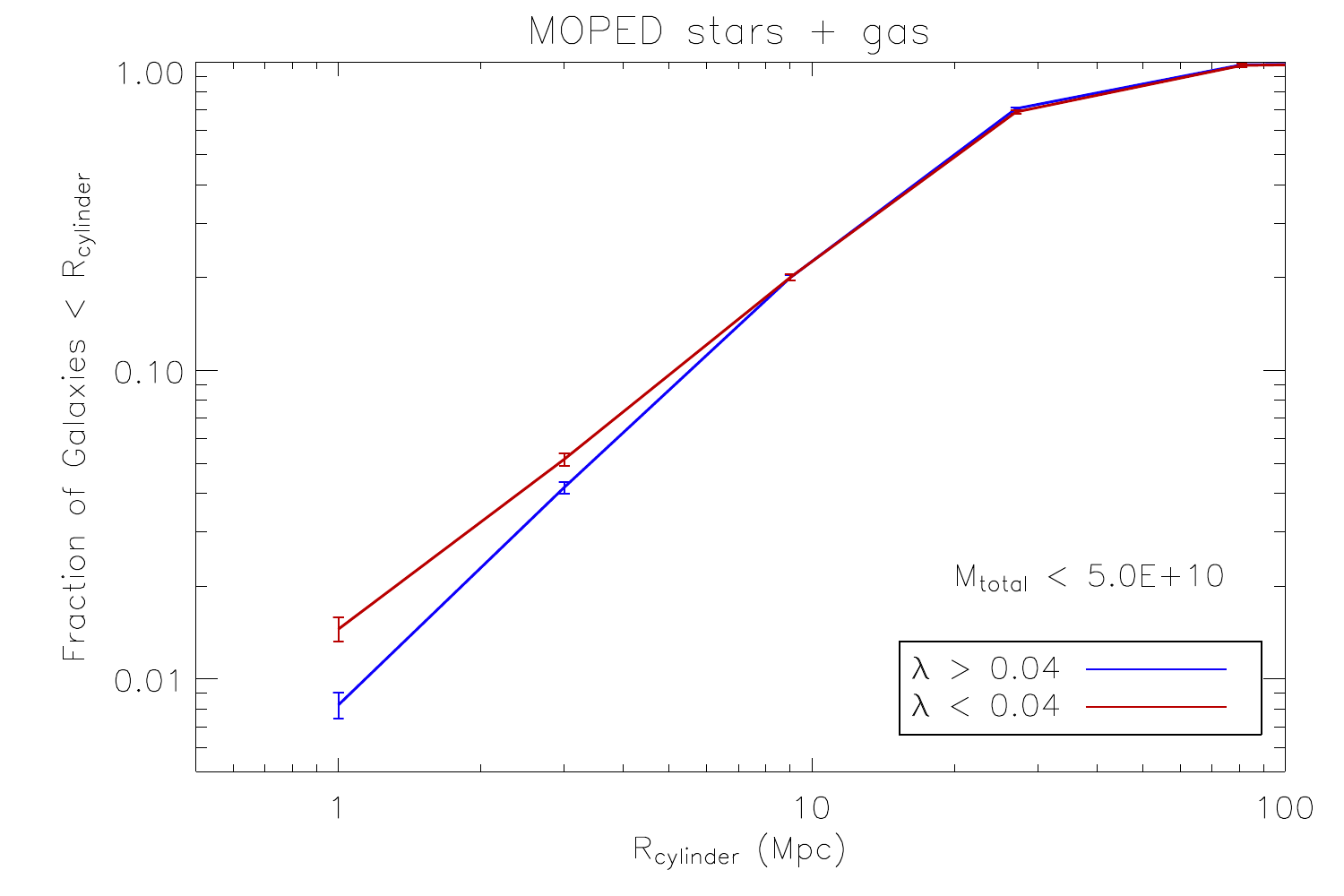}
\includegraphics[width=0.68\columnwidth]{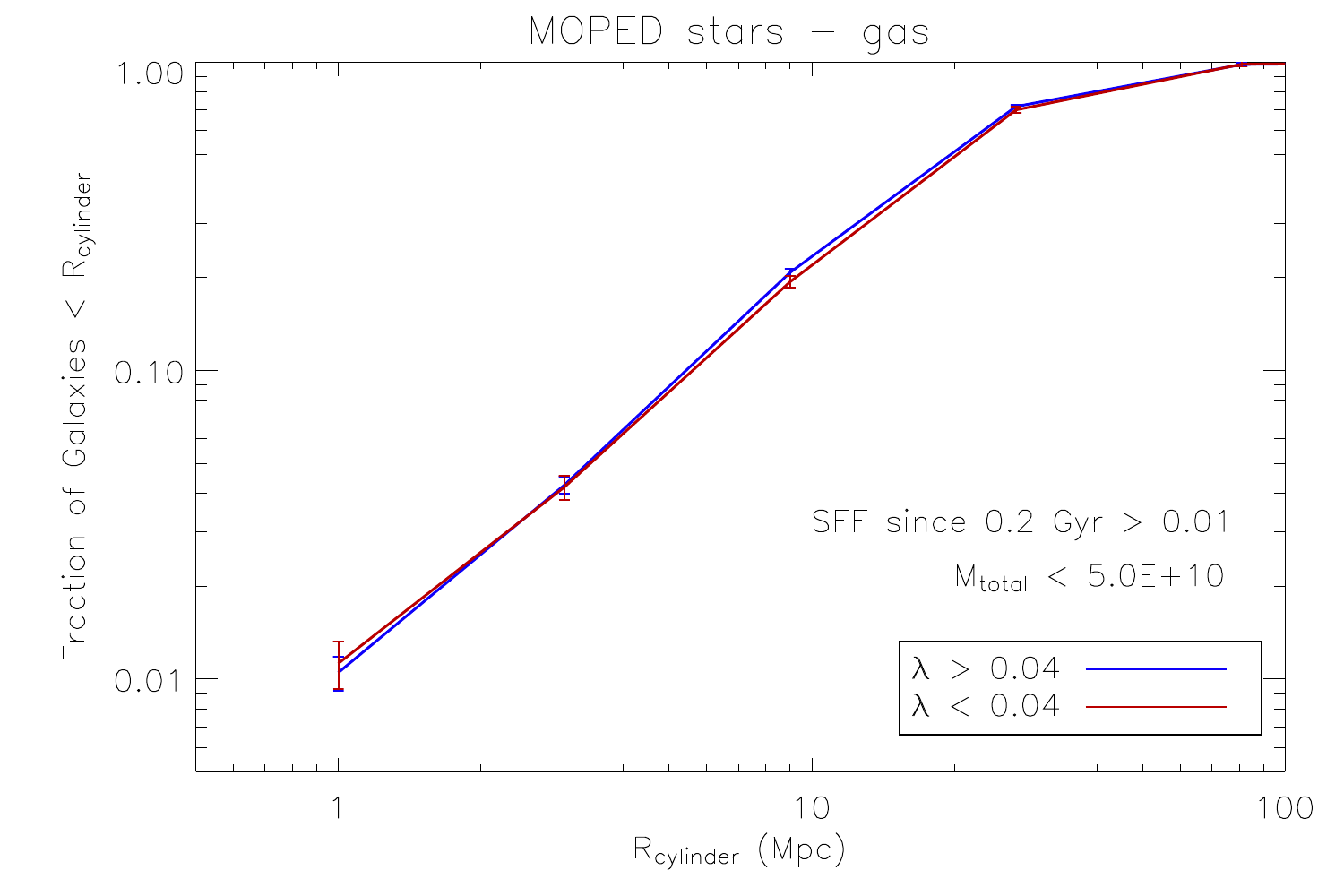} 

\caption{In each panel, the fraction of galaxies for which at least one RMF galaxy cluster is found within a cylinder of depth $\Delta z = 0.02$ as  a function of the radius of that cylinder $R_{cylinder}$. High spin ($\lambda > 0.04$, {\it blue lines}) and low spin ($\lambda < 0.04$, {\it red lines}) are plotted for the entire sample ({\it left}), low-mass galaxies ({\it center}), and low-mass, star-forming galaxies ({\it right}). Poisson errors are shown as lower limits on the true error. Note that these plots are cumulative; the individual points are not independent.}
\label{fig:cluster_radii}
\end{figure*}

\subsection{Enviromental dependence}

We will now investigate the role of environment in determining spin. With the largest sample of (indirectly measured) spins yet collected, we are well positioned to answer this question. We approach this question in two ways: first we compare to a sample of galaxy clusters found in the SDSS-DR6 catalogue using a new algorithm (Berta 2008, in preparation), and second we use a marked correlation study \citep{marks06} to compute the clustering properties of the spin itself.

The cluster catalogue was recently constructed from DR6 galaxy photometry with the Rosy Matched Filter (RMF) cluster-finding algorithm, which integrates a Matched Filter technique \citep{Postman96, Kepner99, Kim02} with an early-type red-sequence search criterion like that used by the Red-Sequence Cluster Survey and the maxBCG cluster catalogue \citep{GladdersYee, Koester}. Clusters are identified as peaks in a 3D likelihood map measuring how well observed galaxies match a realistic cluster filter in a 5D position-brightness-color-color-color space, and characterized by a richness parameter $\Lambda_{RMF}$, designed to scale with total cluster luminosity. Note, the RMF catalogue was compiled only from bulge-dominated galaxies with {\tt fracDev > 0.5}, a completely disparate set from the galaxies studied here.

For each galaxy for which we measure a spin, we calculate a radius $R_{cylinder}$ such that an imaginary cylinder of depth $\Delta z = 0.02$ and radius $R_{cylinder}$ is just large enough to include the one RMF cluster. {This is basically the projected distance to the nearest cluster, but it is expresed in this manner} because precise galaxy-to-cluster distances are nearly impossible to measure (Finger-of-God effects, cluster redshift errors, etc...). We will use $R_{cylinder}$ as a rough tracer of the local density near each galaxy, with smaller $R_{cylinder}$ corresponding to denser environments. Fig.~\ref{fig:cluster_radii} shows the cumulative distribution of $R_{cylinder}$ for high and low spins, with each curve being normalized to its last bin. In these plots, shallower slopes correspond to more clustered galaxies.

The first panel of Fig.~\ref{fig:cluster_radii} seems to indicate that low-$\lambda$ galaxies are much more strongly clustered than high-$\lambda$ galaxies, {and such a finding is in accordance with \citet{avila}}, but is this difference a result of the spins alone? It has long been known that clusters are overly populated by massive, early-type galaxies with old stellar populations. As indicated above, these galaxies are more likely to sit in low-$\lambda$ halos and could thus contribute to the observed clustering difference. The second panel shows only low-mass galaxies ($M_{total} < 5\times10^{10}$ M$_{\odot}$, roughly the median of the complete sample), and as expected, the difference in clustering strengths decreases, although still indicating low-$\lambda$ galaxies to lie preferentially nearer to clusters. In the third panel, we include only the low-mass ($M_{total} < 5\times10^{10}$ M$_{\odot}$) galaxies which have formed more than 1\% of their stars since 0.2 Gyr (this corresponds to top panel of Fig.~\ref{fig:sfspin}), and all visible clustering dependence on spin vanishes. It seems the observed $R_{cylinder}-\lambda$ correlation can be adequately explained as the confluence of galaxy mass and stellar age increasing towards cluster centers.

\begin{figure}
\includegraphics[width=\columnwidth]{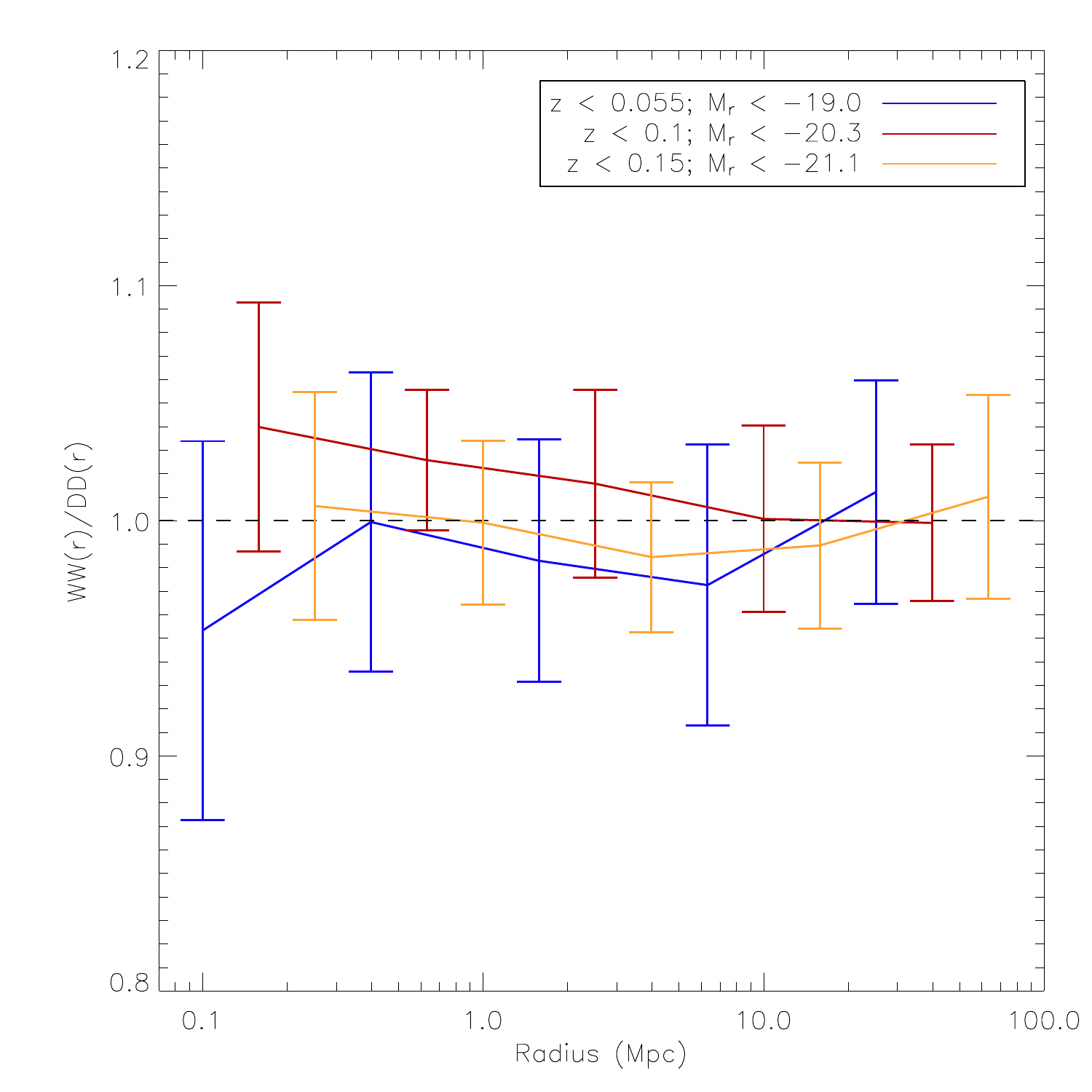}
\caption{A marked correlation function showing the $\lambda$-weighted galaxy pair counts as a function of separation radius $WW(r)$ normalized by the unweighted pair counts $DD(r)$. Three volume limited samples are shown ($z<0.055$, $z<0.1$, $z<0.15$). Error bars are calculated from 6 jacknife realizations of  $WW(r)/DD(r)$.}
\label{fig:mark_corr}
\end{figure}

{Using RMF clusters as density tracers has the advantage of covering a wide volume and broad range of galaxy masses, but to} investigate further the correlation properties of the dark halo spin we performed a marked correlation analysis \citep{marks06} {in several volume limited samples}, using galaxy spin $\lambda$ as the mark. Fig.~\ref{fig:mark_corr} shows the deviation of $\lambda$ from its mean value as a function of scale, as calculated in three volume limited samples: roughly the HC-S06 volume $z<0.055$, roughly the ``faint'' volume from \citet{marks06} $z<0.1$, and a deeper volume $z<0.15$, {using the {\tt MOPED stars + gas} estimate in all cases.} The associated absolute magnitude cuts are shown in Fig.~\ref{fig:mark_corr}. 

We find no significant scale dependence of $\lambda$, {a result which persists among the {\tt BTF}, {\tt MOPED stars}, and {\tt MOPED stars + gas} estimates. Each volume corresponds to a much narrower range of galaxy masses than Fig.~\ref{fig:cluster_radii}, which accounts for the absence of any such clustering dependence as seen in the first panel of Fig.!\ref{fig:cluster_radii}.}

This is what is predicted in analytical and numerical results \citep{HP88,Gouda98,bett} but in contrast with the results of \citet{falten} who explicitly computed the marked correlation of the spin parameter $\lambda$ in their simulations and found an environmental dependence. However, \citet{falten} argue that large samples are needed to measure this signal, it therefore remains to be seen if much larger sample can confirm our result. \citet{Hernandez08} also recently found no environmental dependence for $\lambda$.

\section{Conclusions}

We have used an indirect method to compute the dark matter spin of galaxies and combined it with MOPED determinations of the star formation history of the SDSS galaxies.   Exploiting the large sample available, we have studied the influence of dark matter spin on galaxy mass, star formation history, and environment.  We find that galaxy dark matter spin and stellar mass are anti-correlated: lower stellar mass galaxies exhibit broader and generally higher distribution of spins than high-mass galaxies. {Furthermore, according to halo mass estimates determined from the {\tt MOPED} stellar masses, galaxies which have formed $1-6\%$ of their stellar mass in the past $0.2$ Gyr also have typically broader and higher-$\lambda$ spin distributions than galaxies that have formed a large fraction of their stellar mass at look-back times larger than $10$ Gyr.} Although mass is the prime parameter determining the current star formation rate, the galaxy spin parameter might play a weak secondary role, with higher-spin galaxies having more current star formation at given stellar mass. 

{Such a result could be explained in terms of high-$\lambda$ halos hosting low surface density disks with consequently less efficient star formation rates in the past leaving a larger reservoir of gas today. This role of spin in galaxy formation was also emphasized by \citet{JHHP97}, who proposed a `spin bias'. We find here that though this effect might exist it is likely a subdominant effect. Along the lines of \citet{Vitvitskaetal2002, Maller&Dekel2002}, the link between star formation and $\lambda$ might follow from high spin galaxies experiencing recent mergers. Even if these mergers are merely the signature of the large scale tidal torques on hierarchical galaxy formation, they might both increase spin and spur starbursts in an individual galaxy, provided the stellar disk survives the merger.}

We have also looked at environmental effects: using the RMF catalogue of galaxy clusters in the SDSS we find a very weak anti-correlation between the value of dark matter spin and proximity to a cluster, but such as would be consistent with mass and star formation being positively correlated with cluster proximity (see \citet{Abilio} for a comparison of the environmental dependence of galaxies with numerical models of galaxy formation). A marked correlation study also shows no strong correlation with galaxy separation, {in agreement with previous studies' predictions.}

{Finally, we note that this analysis requires an estimation of three unobservable galaxy parameters $M$, $J$, and $E$, which is not at all a trivial problem. Determining the latter two can be difficult even in numerical simulations: although each dark matter particle can be followed, arbitrarily defined galaxy borders can result in arbitrary results for $\lambda$ \citep{DOnghia&Navarro}. Furthermore, as the link between the observed stellar disk and the total halo angular momentum is complicated \citep[ie]{vandenBoschetal2002}, the problem is very difficult. We used three estimates for the halo mass $M_{h}\approx M$, and found that they did change our results somewhat. 

Although {\tt MOPED} probably gives a much more accurately measure of the true baryonic mass of a galaxy than the double-scaling relation {\tt BTF}, further study is necessary to elucidate the accuracy of the estimator used here.}

\section*{acknowledgements}
The research of RJ is partially supported by NSF grant PIRE-0507768. We acknowledge funding from the Spanish Ministerio de Educacion y Ciencia (MEC), the CSIC and the European Union FP7 program. We warmly thank David Spergel and Licia Verde for useful comments on the manuscript.

Funding for the creation and distribution of the SDSS Archive has been provided by the Alfred P. Sloan Foundation, the Participating Institutions, the National Aeronautics and Space Administration, the National Science Foundation, the U.S. Department of Energy, the Japanese Monbukagakusho, and the Max Planck Society. The SDSS Web site is {\tt http://www.sdss.org/}.

The SDSS is managed by the Astrophysical Research Consortium (ARC) for the Participating Institutions. The Participating Institutions are The University of Chicago, Fermilab, the Institute for Advanced Study, the Japan Participation Group, The Johns Hopkins University, the Korean Scientist Group, Los Alamos National Laboratory, the Max-Planck-Institute for Astronomy (MPIA), the Max-Planck-Institute for Astrophysics (MPA), New Mexico State University, University of Pittsburgh, University of Portsmouth, Princeton University, the United States Naval Observatory, and the University of Washington.

\end{document}